\documentclass[acs,groupedaddress,twocolumn]{revtex4}

\usepackage{amsmath}
\usepackage{graphicx}
\usepackage{dcolumn}
\usepackage{bm}
\usepackage{alltt}%

\begin{document}

\title{Free energy calculations: An efficient adaptive biasing potential method}

\author{Bradley M. Dickson}
\email{adynata@gmail.com}
\affiliation{Universit\'e Paris Est, CERMICS, Project MICMAC Ecole des Ponts ParisTech - INRIA, 6 \& 8 Av. Pascal, 77455 Marne-la-Vall\'ee Cedex 2, France}
\affiliation{Laboratoire de Chimie, \'{E}cole Normale Sup\'{e}rieure de Lyon, 46 all\'{e}e d'Italie, 69364 Lyon, France}
\author{Fr\'{e}d\'{e}ric Legoll}
\affiliation{Universit\'e Paris Est, Institut Navier, LAMI, Project MICMAC Ecole des Ponts ParisTech - INRIA, 6 \& 8 Av. Pascal, 77455 Marne-la-Vall\'ee Cedex 2, France}
\author{Tony Leli\`{e}vre}
\affiliation{Universit\'e Paris Est, CERMICS, Project MICMAC Ecole des Ponts ParisTech - INRIA, 6 \& 8 Av. Pascal, 77455 Marne-la-Vall\'ee Cedex 2, France}
\author{Gabriel Stoltz}
\affiliation{Universit\'e Paris Est, CERMICS, Project MICMAC Ecole des Ponts ParisTech - INRIA, 6 \& 8 Av. Pascal, 77455 Marne-la-Vall\'ee Cedex 2, France}
\author{Paul Fleurat-Lessard}
\affiliation{Laboratoire de Chimie, \'{E}cole Normale Sup\'{e}rieure de Lyon, 46 all\'{e}e d'Italie, 69364 Lyon, France}

\date{\today}

\begin{abstract}
We develop an efficient sampling and free energy calculation technique
within the adaptive biasing potential (ABP) framework.  By mollifying
the density of states we obtain an approximate free energy and an
adaptive bias potential that is computed directly from the population
along the coordinates of the free energy. Because of the mollifier,
the bias potential is ``nonlocal'' and its gradient admits a simple 
analytic expression. 
A single observation of the reaction
coordinate can thus be used to update the approximate free energy at
every point within a neighborhood of the observation.  This greatly
reduces the equilibration time of the adaptive bias potential.  This
approximation introduces two parameters: strength of mollification and
the zero of energy of the bias potential.  While we observe that the
approximate free energy is a very good estimate of the actual free
energy for a large range of mollification strength, we demonstrate that the
errors associated with the mollification may be removed via
deconvolution.  The zero of energy of the bias potential, which is
easy to choose, influences the speed of convergence but not the
limiting accuracy.  This method is simple to apply to free energy or
mean force computation in multiple dimensions and does not involve
second derivatives of the reaction coordinates, matrix manipulations
nor on-the-fly adaptation of parameters.  For the alanine dipeptide
test case, the new method is found to gain as much as a factor of ten in
efficiency as compared to two common adaptive biasing force
formulations and it is shown to be as efficient as well-tempered metadynamics 
with the post-process deconvolution giving a clear advantage to the mollified density of states method. 
\end{abstract}

\maketitle

\section{Introduction}

Many interesting physical systems can be categorized as rare-event
systems. The uniting feature of these systems is that the dynamics
involved require a time resolution much smaller than the timescale on
which interesting events take place. Of central importance to the
evolution of such systems is the free energy. Low lying regions of the
free energy and the barriers separating these regions dictate the
thermodynamics and, to some extent, the kinetics\cite{ht90} of the
system. Because free energy barriers are only rarely crossed,
efficient exploration of the free energy landscape is practically
impossible with straightforward integration of the equations of
motion.

Recently, a number of approaches have used the history of the dynamics
to accelerate exploration of the free energy
landscape\cite{wl01,dp01,lp02,mb06,mv06,lrs07}.  In these methods
information about the free energy is estimated during simulation and
that information is fed back to the dynamics as a statistical
bias. While there are many variations on this idea, the common aim is
to minimize the time spent sampling regions of the free energy that
have been sampled in the past.  These schemes may be classified into
two categories: adaptive bias force methods (ABF)\cite{dp01,dp08}
which use an approximation of the mean force to bias the dynamics;
adaptive biasing potential methods (ABP)\cite{wl101,lp02,mb06} which
use an approximation of the free energy as a bias potential.

The underlying idea for all adaptive methods is that it is 
computationally more efficient to sample the distribution associated with a
flattened free energy than it is to sample the density associated with
the actual, very rough free energy.  We propose an ABP method that builds
an approximate density of states (DOS) and uses that approximation to 
define a bias potential.  
Mollification of the underlying density of states produces the
desired approximation and leads to a smooth, adaptive bias potential 
whose gradient admits a simple analytic expression 
and that can be computed without knowledge of the actual density of states. 
Because
the actual and approximate free energies are related by a convolution,
it is easy to recover the former from the latter via deconvolution.
Our framework is not restricted to one-dimensional or orthogonal 
reaction coordinates. Moreover, it 
avoids second derivatives of the reaction
coordinate.

This paper is organized as follows. We describe our ABP method in 
Section~\ref{sec:description} (see in particular 
Eqs.~\eqref{Cbias}, \eqref{CbFE} and~\eqref{CbwMD}), and comment on its convergence.
We contrast this method to existing ones in Section~\ref{sec:comparison}, 
and present some numerical validation on a benchmark system in 
Section~\ref{sec:numerical}. Our conclusions are summarized in Section~\ref{sec:ccl}.

\section{Description of the method}
\label{sec:description}

\subsection{Free energy and its mollified version}

Consider a system whose configuration is described
by a variable $x \in {\mathcal{X}}$, where ${\mathcal{X}}$ is the configuration
space. We denote by $V$ the potential energy function.
Assume that we are given a $N$-dimensional reaction coordinate 
$\xi(x)$ with values in $\Omega$, which characterizes some physical event.
The density of states at a value $\xi^*$ of the reaction coordinate is
defined as
\begin{equation}
  \label{eq:def_DOS}
  e^{-\beta A(\xi^*)} = Z^{-1} \int_{\mathcal{X}} \delta(\xi(x)-\xi^*) \, e^{-\beta V(x)} \, dx,
\end{equation}
where $\beta = 1/(k_{B}T)$, and $Z$ is a normalization constant, chosen such that
\[
\int_\Omega e^{-\beta A(\xi)} \, d\xi = 1.
\]
Eq.~\eqref{eq:def_DOS} defines the free energy $A(\xi^*)$. Recall that, in practice,
the free energy needs only be known up to an additive constant since the important quantities
to describe the relative likelihoods of physical states are free-energy differences.

In general, the free energy is unknown and has to be approximated.
The method we propose in this work is based on the following limit:
\[
e^{-\beta A(\xi^*)} = \lim_{\alpha \rightarrow 0} e^{-\beta A_\alpha(\xi^*)},
\]
where
\begin{equation}
  \label{eq:approx_FE}
  e^{-\beta A_\alpha(\xi^*)} = 
  Z^{-1} \int_{\mathcal{X}} \delta_{\alpha}(\xi(x)-\xi^*) \, e^{-\beta V(x)} \, dx, 
\end{equation}
with for instance a Gaussian approximation of the
Dirac delta function:
\[
\delta_{\alpha}(\xi) = \left(\frac{1}{\alpha \sqrt{\pi}} \right)^N 
\exp\left(-\frac{|\xi|^2}{\alpha^2}\right).
\]
Equation~\eqref{eq:approx_FE} defines an approximate free energy $A_{\alpha}$,
obtained by sampling the density of states at a finite $\alpha$,
\textit{i.e.}, by sampling a mollified density of states. 
Notice that the approximation resulting from finite $\alpha$ 
can in fact be rewritten as a convolution of the actual density of 
states $e^{-\beta A}$ with $\delta_{\alpha}$. Indeed,
\begin{eqnarray}
\label{conv}
&& e^{-\beta A_{\alpha}(\xi^*)} \\
&&= Z^{-1}\int_{\mathcal{X}} \delta_{\alpha}(\xi(x)-\xi^*) \, e^{-\beta V(x)} \, dx \nonumber \\
&& = Z^{-1}\int_\Omega \int_{\mathcal{X}} \delta_{\alpha}(\bar{\xi}-\xi^*) \, 
\delta(\xi(x)-\bar{\xi}) \, e^{-\beta V(x)} \, dx \, d\bar{\xi} \nonumber\\
&& = \int_\Omega \delta_{\alpha}(\bar{\xi}-\xi^*) \, e^{-\beta A(\bar{\xi})} 
\, d\bar{\xi}.\nonumber
\end{eqnarray} 
This remark is the basis for an extraction of the actual 
free energy $A$ from $A_{\alpha}$ through a deconvolution procedure 
(see Section~\ref{sec:simu_details}). 
While we make this presentation with a scalar $\alpha$, 
this could easily be generalized to the case where $\alpha$ takes 
different values in different dimensions of the reaction coordinate. 

Equation~\eqref{conv} is also helpful in assessing the errors introduced 
in umbrella sampling (US) and thermodynamic integration (TI) simulations 
employing harmonic constraint potentials.
The corresponding error is analogous to the
convolution errors discussed in this paper.  
Note that the parameter $\alpha$ can be converted to a force
constant for a harmonic potential via $k = 2k_BT/\alpha^2$, where $k$
is the force constant. Errors resulting from finite $k$ in TI and US
computations can be identified as resulting from a convolution between
the true density of states and a known Gaussian function.  Typically,
the harmonic constraints are tight enough for $A_{\alpha}$ to be a
good approximation of $A$ but any persisting bias can, at least in
principle, be removed by deconvolution as shown below.

\subsection{Interest of the mollified free energy}

In this work, we
use $A_{\alpha}$ to define an adaptive bias. 
The first interest of this approach is that the gradient of $A_{\alpha}$ is much easier to 
compute than the gradient of $A$. Indeed, the laster reads (see References~[\citenum{o98,sc98,cle07}])
\begin{equation}
  \label{cle}
  F_j(\xi^*) = \left\langle \displaystyle\sum_{i=1}^N \nabla V \cdot G^{-1}_{ji}\nabla \xi_{i}
  - \beta^{-1}\nabla \cdot (G^{-1}_{ji}\nabla \xi_{i}) \right\rangle_{\xi^*},
\end{equation}
where $\langle \cdot \rangle_{\xi^*}$ denotes a canonical average for a fixed value of the 
reaction coordinate, and $G$ is the Gram matrix. The latter matrix is defined
as $G = J J^t$ with $J_{ij} = \partial \xi_i/\partial x_j$ ($x_i$ are the Cartesian coordinates on which
the reaction coordinates are defined).  
The computation of the free energy gradient therefore requires the computation
of second derivatives of the reaction coordinate, which is cumbersome in many cases.
The gradient of the mollified free energy has a much simpler expression:
\begin{equation}
  \label{convFg}
  \frac{\partial A_{\alpha}(\xi^*)}{\partial \xi^*_j} = -k_B T\frac{\displaystyle \int_{\mathcal{X}}
    \partial_{\xi^*_j} \delta_{\alpha}(\xi(x)-\xi^*) \, e^{-\beta V(x)} \, dx}
       {\displaystyle \int_{\mathcal{X}} \delta_{\alpha}(\xi(x)-\xi^*) \, e^{-\beta V(x)} \, dx},
\end{equation} 
where $j$ is a reaction coordinate index and
\[
\partial_{\xi^*_j} \delta_{\alpha}(\xi_j(x)-\xi^*) =
\frac{2}{\alpha^2} (\xi_j(x)-\xi_j^*) \delta_{\alpha}(\xi(x)-\xi^*).
\]
In particular, no derivative of the reaction coordinates are required.

Another interest of the mollified free energy
lies in the nonlocality of $\delta_{\alpha}$, which allows a single observation of
$\xi$ to contribute to $A_\alpha$ for a range of values $\xi^*$,
leading to a faster convergence. The
question is then whether there is a range of $\alpha$ for which: (i)
$\alpha$ is sufficiently large so that $A_\alpha$ could be estimated
with fewer samples than what would be required to compute $A$ and (ii)
$\alpha$ is sufficiently small, so that $A_\alpha$ is close enough to
$A$ to efficiently bias the dynamics. We show in Section~\ref{sec:choice_alpha}
that a large range of $\alpha$ satisfies these two conditions 
on a paradigmatic test case.

\subsection{A new ABP method}

\subsubsection{Construction of the method}

To compute approximations of~\eqref{conv} and~\eqref{convFg} as time averages along
a trajectory $x_t$ driven by the potential function $V(x)$, we first
assume that $x_t$ is ergodic with respect to the canonical ensemble.
We may take $x_t$ as a solution to the Langevin equation driven by the potential $V$, 
for example. Using trajectory averages, \eqref{conv} can be obtained as the following 
longtime limit:
\begin{equation}
  \label{CweakFE}
  e^{-\beta A_{\alpha}(\xi^*,t)} 
  = Z_t^{-1} \left(1+\displaystyle\int_0^t \delta_{\alpha}(\xi(x_s)-\xi^*) \, ds\right), 
\end{equation} 
where the normalization constant $Z_t$ is
\[
Z_t = \int_\Omega \left(1+\displaystyle\int_0^t \delta_{\alpha}(\xi(x_s)-\xi^*) \, ds\right)
d\xi = |\Omega|  + t.
\]
The normalization constant ensures that
\begin{equation}
  \label{eq:normalization_mollified}
  \int_\Omega e^{-\beta A_{\alpha}(\xi,t)} \, d\xi = 1
\end{equation}
at all times $t \geq 0$.
Notice that we implicitely assumed that the reaction coordinate has values
in a finite space $\Omega$. This is indeed the case when angles are considered.
For unbounded reaction coordinates, it is always possible to restrict the sampling
to important values of $\xi(x)$. In practice, the range of the reaction
coordinate needs to be truncated anyway.

From~\eqref{CweakFE}, we obtain 
\begin{equation}
  \label{CweakFEg}
  \frac{\partial A_{\alpha}(\xi^*,t)}{\partial \xi^*_j} = 
  -k_B T\frac{\displaystyle\int_0^t \partial_{\xi^*_j}\delta_{\alpha}(\xi(x_s)-\xi^*) \, ds}
  {\displaystyle 1+\int_0^t \delta_{\alpha}(\xi(x_s)-\xi^*) \, ds}, 
\end{equation} which, in the longtime limit converges to~\eqref{convFg}

Now, a simple ergodic average such as \eqref{CweakFE} or \eqref{CweakFEg} 
can of course not be used
in practice since the dynamics at hand are usually metastable for complex systems,
and the convergence of the time averages~\eqref{CweakFE} and~\eqref{CweakFEg} is very slow.
We therefore need to bias the dynamics in order to remove the metastability.

In what follows, we will consider a trajectory $x_t$ obtained from the equations of motion 
with the biased potential $V+V_b$. 
The idea behind adaptive method is to use the opposite of some current approximation 
of the free energy as a biasing potential, and to update the estimate
as time goes on, in a way such that the bias eventually converges
to the correct free energy.
Here, we consider an adaptive biasing potential method,
defined through the following update of the biasing potential $V_b$:
\begin{equation}
  \label{Cbias}
  e^{\beta V_b(\xi,t)} = e^{-\beta \Delta A_{\alpha}(\xi,t)} \, e^{\beta c}
\end{equation}
where the renormalized current approximation of the mollified free energy
$e^{-\beta \Delta A_{\alpha}(\xi,t)}$ is
\[
e^{-\beta \Delta A_{\alpha}(\xi,t)} = 
\frac{e^{-\beta A_{\alpha}(\xi,t)}}{\displaystyle \max_{\xi^*}\left[e^{-\beta A_{\alpha}(\xi^*,t)}\right]}. 
\]
The parameter $c$ in~\eqref{Cbias} is an important quantity in our method, which allows
to tune the convergence rate of the method. We discuss its choice in Section~\ref{sec:cv}. 
With these definitions, $V_b = -A_\alpha$ up to an additive constant which is chosen such that 
$\max[V_b] = c$. Similarly, $\Delta A_\alpha = A_\alpha$, again, up to an additive constant which 
is such that $\min[\Delta A_\alpha] = 0$
 
Departing from standard ABP/ABF frameworks 
we use ideas from importance sampling to write 
\eqref{CweakFE} and \eqref{CweakFEg} as time averages over biased trajectories 
\begin{equation}
  \label{CbFE}
  e^{-\beta A_{\alpha}(\xi^*,t)} 
  = Z_t^{-1} \left(1+\displaystyle\int_0^t \delta_{\alpha}(\xi(x_s)-\xi^*) \, 
  e^{\beta V_b(\xi(x_s),s)}\, ds\right), 
\end{equation} 
where $Z_t$ is still a normalization constant 
ensuring~\eqref{eq:normalization_mollified}, and
\begin{equation}
  \label{CbwMD}
  \frac{\partial A_{\alpha}(\xi^*,t)}{\partial \xi^*_j} = 
  -k_B T\frac{\displaystyle\int_0^t \partial_{\xi^*_j}\delta_{\alpha}(\xi(x_s)-\xi^*) \, 
    e^{\beta V_b(\xi(x_s),s)}\, ds}
  {\displaystyle 1+\int_0^t \delta_{\alpha}(\xi(x_s)-\xi^*) \, e^{\beta V_b(\xi(x_s),s)} \, ds}. 
\end{equation} 
The ABP method we discuss here is based on the biasing potential~\eqref{Cbias},
updated with the current estimate of the free energy~\eqref{CbFE}. New configurations are
obtained by integrating in time the biased equations of motion using the simple
estimate~\eqref{CbwMD} for the biasing force.
The convergence of this method is discussed in Section~\ref{sec:cv}.

In fact, Eq.~\eqref{CbFE} is a way to evaluate the convolution in
Eq.~\eqref{conv} at each point $\xi^*$ using a biased trajectory. This
gives us a precise understanding of how using finite $\alpha$
introduces error in the estimate $A_{\alpha}$ and how to remove that
error. This is a strength of our method which makes it unique.  If we
try to draw analogy with metadynamics, the framework 
of~\eqref{CbFE} would imply the continuous deposition of the Gaussians
$\delta_{\alpha}$ at each point $\xi(x_t)$ along the trajectory.
Notice that in this analogy the Gaussians would be added to the
density of states rather than to the bias potential, precluding us
from going any further with the analogy.

\subsubsection{Time-discretization}

Let us briefly discuss the time-discretization of the method
based on \eqref{Cbias}-\eqref{CbFE}-\eqref{CbwMD}.
Assume that we have a suitable discretization where time is broken into
parts of duration $\Delta t$ so $t = n\Delta t$ and $x_{i\Delta t}$ is
written $x_i$. The biasing potential is now updated as
\begin{equation}
  \label{bias}
  e^{\beta V_b(\xi,n)} = e^{-\beta \Delta A_{\alpha}(\xi,n)}e^{\beta c}
\end{equation}
where $e^{-\beta \Delta A_{\alpha}(\xi,n)} = 
e^{-\beta A_{\alpha}(\xi,n)}/\max_{\xi^*}[e^{-\beta A_{\alpha}(\xi^*,n)}]$, and 
\eqref{CbFE} and \eqref{CbwMD} are respectively replaced by  
\begin{eqnarray}
  \label{bFE}
  && e^{-\beta A_{\alpha}(\xi^*,n+1)} \\
  && = Z^{-1}_n 
  \left(1+\displaystyle\sum_{i=0}^n \delta_{\alpha}(\xi(x_i)-\xi^*)
  \, e^{\beta V_b(\xi(x_i),i)}\right),
  \nonumber
\end{eqnarray} 
and
\begin{eqnarray}
  \label{bwMD}
  && \frac{\partial A_{\alpha}(\xi^*,n+1)}{\partial \xi^*_j} \\
  && = -k_BT
  \frac{\displaystyle\sum_{i=0}^n \partial_{\xi^*_j}\delta_{\alpha}(\xi(x_i)-\xi^*) 
    \, e^{\beta V_b(\xi(x_i),i)}}{1+\displaystyle\sum_{i=0}^n \delta_{\alpha}
    (\xi(x_i)-\xi^*) \, e^{\beta V_b(\xi(x_i),i)} }. \nonumber
\end{eqnarray} 
At $t=0$ we have $\exp[-\beta A_{\alpha}(\xi,0)] = 1/Z_0$. 
Let us emphasize again that the trajectory $x_i$ is generated from biased equation of 
motion associated with the biased potential $V+V_b$. 

The implementation only requires storing the current value of 
the numerator and denominator of Eq.~\eqref{bwMD} at the points $\xi^*$. In particular, $Z_n$ is
never needed in practice. 
(see Appendix \ref{app:algo}) 
The biasing force $-\nabla V_b$, needed for instance to integrate the Langevin dynamics, 
is obtained through Eq.~\eqref{bwMD}.
\subsubsection{Convergence and consistency}
\label{sec:cv}

It can be checked that, if the biasing potential $V_b$ converges 
in the long-time limit, then
it converges to $-A_{\alpha}$ up to an additive constant. 
Indeed, denoting by $\overline{A}_\alpha(\xi) = \lim_{n\to+\infty} A_\alpha(\xi,t)$,
the trajectory $x_i$ is sampled according to the limiting canonical
measure associated with the potential $V-\overline{A}_\alpha + C$ (where $C$ is
an unimportant constant), so that~\eqref{CbFE} leads to
\begin{eqnarray*}
&& e^{-\beta \overline{A}_\alpha(\xi^*)} \\
&& = \lim_{t\to+\infty} 
\frac{1+\displaystyle\int_0^t \delta_{\alpha}(\xi(x_s)-\xi^*) \, e^{\beta V_b(\xi(x_s),s)} \, ds}
{\displaystyle\int_\Omega \left(1+\int_0^t \delta_{\alpha}(\xi(x_s)-\xi') \, 
  e^{\beta V_b(\xi(x_s),s)} \, ds \right) d\xi'}\\
&& =\frac{\displaystyle \int_{\mathcal{X}} \delta_\alpha(\xi(x)-\xi^*) \, e^{-\beta (V(x)+C)} \, dx}
{\displaystyle \int_\Omega \int_{\mathcal{X}} \delta_\alpha(\xi(x)-\xi') \, 
  e^{-\beta (V(x)+C)} \, dx \, d\xi'}\\
&& = e^{-\beta A_\alpha(\xi^*)}.
\end{eqnarray*}
The fact that, if a limit exists, then it is the correct one,  
is an important consistency
check of the method. However, we were not able to prove that the biasing potential
indeed converges (this issue arises in all ABP methods while 
such an alalysis can rigorously be done for some ABF methods\cite{lrs08}).

Let us now look more carefully at the first iterations of the algorithm,
in order to understand the role of the constant $c$ in~\eqref{Cbias} or~\eqref{bias}. 
We base our considerations on the numerical discretization~\eqref{bFE} 
to simplify the argument. First, recall that the constant $c$
does not change the longtime limit of the algorithm.
However, it helps accelerating the convergence during the initial transient regime.
The first iteration of~\eqref{bFE} indeed shows that 
\[
e^{\beta V_b(\xi^*,1)} = e^{\beta c} \frac{1 + \delta_\alpha(\xi(x_0)-\xi^*)
  \, e^{\beta c}}{1 + \delta_\alpha(0) \, e^{\beta c}}.
\]
When $c$ is such
that $e^{\beta c}$ is small, $V_b(\xi^*,1)$ is raised by a small
amount and the gradient of $V_b$ encourages trajectories to move away
from $\xi^*$ to some small extent. By increasing the value of $c$, we
obtain a bias potential that pushes trajectories away from $\xi^*$
more strongly, hence increasing the efficiency of the bias potential, in
particular at the early stages of the process.
We therefore conclude that the value of $c$ should be as large 
as possible while maintaining numerical stability.
Not all ABP methods update their biases according to this rule,
see the comparison between our approach and the standard Self-healing umbrella sampling
algorithm in Section~\ref{sec:SHUS}.

\section{Comparison with other methods}
\label{sec:comparison}

\subsection{Self-healing Umbrella sampling}
\label{sec:SHUS}

Self-healing umbrella sampling\cite{mb06} (SHUS) can be seen as a
special case of the method presented here. SHUS can be written in
terms of Eq.~\eqref{bias} using the following 
time-dependent constant:
\[
e^{\beta c(n)} = \displaystyle \max_{\xi^*}\left[e^{-\beta A_{\alpha}(\xi^*,n)}\right]. 
\]
With this choice we have, $V_b = -A_\alpha$ and $  \int_\Omega e^{\beta V_b(\xi,t)} \, d\xi = 1$. 
This choice for $c(n)$ was suboptimal since the analysis of section \ref{sec:cv} shows that the 
value of $c$ should be as large as possible. Notice also that when the reaction coordinate space 
is discretized into a finite number of bins, the normalization condition 
\eqref{eq:normalization_mollified} should be restated as a sum over bin indexes and the maximal 
value of $\exp[-\beta A_{\alpha}(\xi,n)]$ is therefore less than one. This corresponds to a 
negative value of $c(n)$. We checked for the testcase considered in Section \ref{sec:numerical} 
that our method outperforms SHUS for precisely this reason.

\subsection{Adaptive biasing force}

We compare numerically our approach to two ABF formulations in
Section~\ref{sec:simu_details}.  ABF is a good reference for
comparison because there are no model parameters to choose. Errors
arise only through time and reaction coordinate discretization.  Two
exact formulations of the free energy gradient are~\eqref{cle} above,
and
\begin{equation}
  \label{dp08t}
  F(\xi^*) = -\left\langle \frac{d}{dt}
  \left(M_{\xi}\frac{d\xi}{dt}\right)\right\rangle_{\xi^*},
\end{equation}  
where $M_\xi^{-1}= J M J^t$ with $M$ the mass matrix and $J$
defined in~\eqref{cle} (see reference~[\citenum{dp08}] for this
second expression). We point out that in practice $F(\xi)$
is approximated by a trajectory average $F(\xi,t)$ which is then used to bias 
the dynamics. 
For further details on the expressions~\eqref{cle} and~\eqref{dp08t} or their
numerical implementation, we refer the reader to the cited works.

With ABF, one must address constructing the free energy from an estimation
of its gradient, the calculated field $F$. While there are specific solutions to this
problem\cite{dp08,mv08,k09} we employ a standard variational
formulation. 
We recast this question as an optimization problem where
the objective function
\begin{equation}\label{obj}
I(u) = \int_\Omega \| F(\xi) - \nabla_{\xi}u \|^2 \, d\xi
\end{equation}
is to be minimized. The corresponding Euler-Lagrange equation is
\begin{equation}\label{peq}
\Delta_{\xi} u(\xi) = \nabla_{\xi} \cdot F,
\end{equation} which is just Poisson's equation, to be supplemented with 
appropriate boundary conditions (depending on the domain $\Omega$). The solution $u(\xi)$ is 
the best representation of the free energy given the vector field
$F(\xi)$. This is solved via finite difference in the present work, but
finite elements (or any Galerkin method) could be used as well.

\subsection{Metadynamics}

Because we have developed a method within the adaptive bias potential
paradigm, we also make a comparison to well-tempered
metadynamics\cite{bbp08}. In this formulation of metadynamics the bias
potential in one dimension is given by 
\begin{equation}\label{metad}
V_b^{\rm meta}(\xi,\tau) = \sum_{t'\leq \tau} G(\xi-\xi_{t'},h(\xi,t'),w),
\end{equation} 
where the functions $G(X,H,W)$ are Gaussians of width $W$ 
and height $H$, centered on $X$. We write $V_b^{\rm meta}$ to indicate 
that this is the bias potential generated by metadynamics. The
Gaussian height in well-tempered metadynamics is both dependent on
time and position along the reaction coordinate $h(\xi,t) = \omega
\exp[-V_{b}^{\rm meta}(\xi,t)/k_B\Delta T]\tau_G$. For details of this version of
metadynamics we refer the reader to reference [\citenum{bbp08}]. We
compare to this particular formulation because it requires less
interaction with the user and a choice of parameter values is given in
the cited reference.

\section{Numerical examples}
\label{sec:numerical}

\subsection{Simulation details and results}
\label{sec:simu_details}

Alanine dipeptide is a familiar system for benchmarking sampling
methods\cite{bk97,mv08,sw08,bbp08,dp08,ss09}.  Here, we employ AMBER
with a half femtosecond timestep, no constraints, solvent effects are
modeled with the generalized Born model and we use the ff94
parameterization. The temperature was maintained at $T=300$K with
Langevin dynamics where the collision frequency is $1$~ps$^{-1}$. We
select the common backbone dihedral angles $(\xi_1,\xi_2) =
(\Phi,\Psi)$ as reaction coordinates.

When discretizing the reaction coordinate, it is common to use a small
bin size to be sure that the free energy is correctly captured.  Here, we
use $300$ bins of width $1.2$ degrees.  We will also consider a bin
width of $3.6$ degrees for Eq.~\eqref{dp08t} to examine the influence
of bin size on ABF.  In practice, for Eqs.~\eqref{bFE} and
\eqref{bwMD}, the current configuration along a trajectory may
contribute only to an $m$ by $m$ grid centered around
$(\xi_1(x_t),\xi_2(x_t))$, which amounts to truncating the range
of the Gaussian function~$\delta_\alpha$. The number of bins $m$ were chosen so that
$\delta'_{\alpha}(\xi-\xi^*)$ is negligible for $\xi^*$ outside this
box. For example, when $\alpha = 5^{\circ}$ we use $m=20$. In practice we neglect 
the normalization $Z_n$ as well as the normalization of $\delta_\alpha$. We give 
a schematic algorithm in appendix \ref{app:algo}.

In simulations with equations \eqref{bFE}, \eqref{cle} and
\eqref{dp08t}, we use a ``ramp function'' 
$R(N_{l,k})=\min[1,N_{l,k}/N_0]$ to scale the biasing force 
(see for instance reference~[\citenum{dp08}]), where $N_{l,k}$ is the population in bin
      $(l,k)$ and the parameter $N_0=10$ was optimized for equations
      \eqref{cle} and \eqref{dp08t}. The ramp function scales the
      biasing force so that the initially noisy observations of the
      force do not induce non equilibrium effects. The biasing force
      for the method presented here is given by Eq.~\eqref{bwMD}. The
      biasing force for the ABF methods are given in equations
      \eqref{cle} and \eqref{dp08t}. The biasing forces (and biasing
      potential) are updated at each timestep.

To study sampling efficiency we use the average difference
\begin{equation}
  \label{error}
  d(t) = \frac{1}{n^2}\sum_{k=1}^n \sum_{l=1}^n \left|A_{\text{ref}}(k,l) -
\hat{A}(k,l,t)\right|
\end{equation} 
between the estimated free energy and a reference to be defined below. 
$n$ is the number of bins in each coordinate, $k$ and $l$ are bin
indices.  For Eq.~\eqref{cle} and \eqref{dp08t} $\hat{A}$ is the
solution of Eq. \eqref{peq}.  
In out method, $\hat{A}$ is either
the left-hand side of Eq.~\eqref{bFE}, $A_{\alpha}$, or its deconvoluted
version $A_{\alpha}^{\rm dcnvl}$. Finally for Eq.~\eqref{metad}, $\hat{A}=-(T+\Delta
T)V_b^{\rm meta}/\Delta T$. The reported results for $d(t)$ are found
by using only a single trajectory with each method.
We do not report the results obtained with SHUS
since the convergence was found to happen much slower than for cases where $c>0$. 

\begin{figure*}
\includegraphics[width=7in]{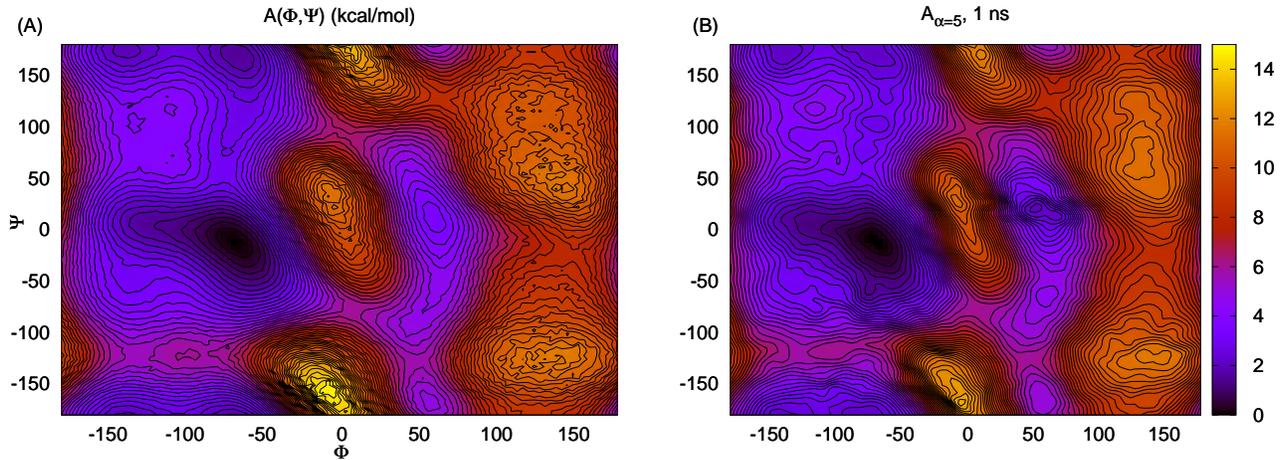}
\caption{Contours are placed every $1/2 k_B T$ (kcal/mol).  (A) an
  estimate of the exact free energy (see text) which compares well to
  reference [\onlinecite{dp08}]. (B) the free energy estimate after 1
  ns of biased dynamics ($\alpha = 5$). \label{ugly}}
\end{figure*}
\begin{figure}
\includegraphics[width=\columnwidth]{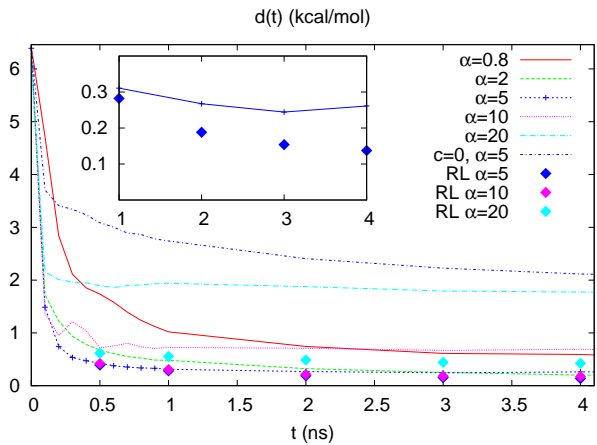}
\caption{Error~\eqref{error} as a function of time for the method presented
  in this paper, for various $\alpha$ with and
  without deconvolution. Unless otherwise stated, $c=15 k_BT$. The
  $\alpha=5$, $c=0$ simulation demonstrates slow convergence due to a
  suboptimal choice of $c$, as described in the text. In the inset we
  show the last 4 ns of the $\alpha = 5^{\circ}$ results.
\label{alpcon}}
\end{figure}
\begin{figure}
\includegraphics[width=\columnwidth]{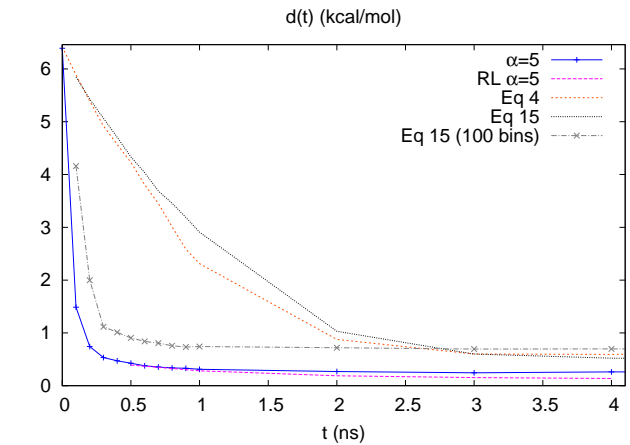}
\caption{Error~\eqref{error} as a function of time for $\alpha=5^{\circ}$ and
  $\alpha=10^{\circ}$ with the method proposed in this paper, 
  and comparison with ABF results obtained from Eqs.~\eqref{cle} and~\eqref{dp08t}.
\label{compcon}}
\end{figure}
\begin{figure*}
\includegraphics[width=7in]{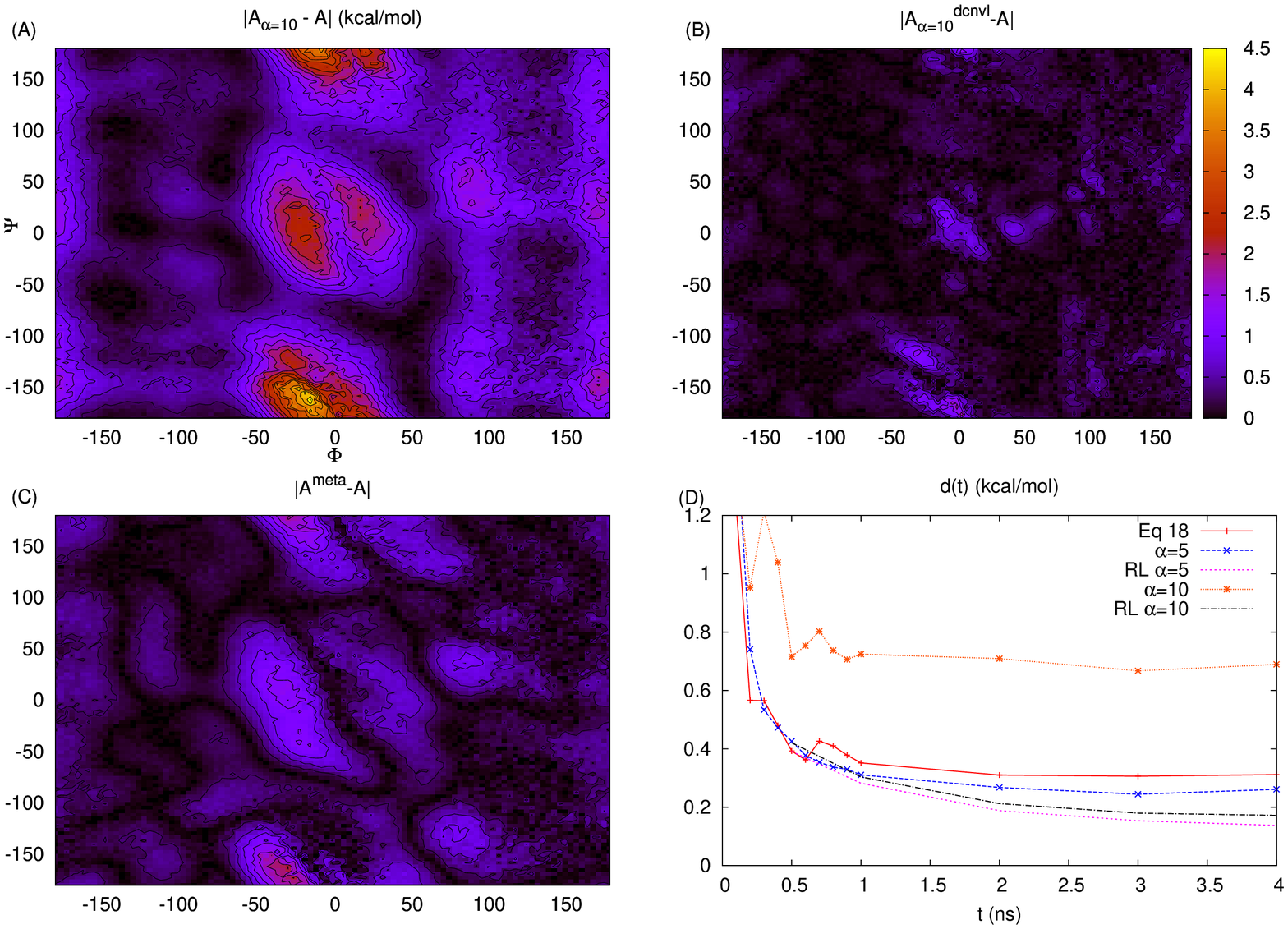}
\caption{In panel (A) we show the absolute difference between the computed 
$A_{\alpha=10}$ and the reference $A$. Most of the error is, as expected, due to the regions of large 
curvature. In panel (B) the absolute difference between the deconvoluted free energy $A_{\alpha=10}^{dcnvl}$ and the reference $A$ 
is shown. In panel (C) we show the absolute difference between $A^{meta} = -(T+\Delta T)V_b^{meta}/\Delta T$ 
and the reference $A$. The 
free energy estimates in panels A-C were taken at the end of a $4$ ns trajectory. 
In panel (D)~\eqref{error} is shown for the well-tempered metadynamics method~\eqref{metad} for comparison 
with results from the mollified DOS method. 
\label{metafig}}
\end{figure*}

We use the Richardson-Lucy algorithm\cite{r72,l74} to deconvolute
$A_{\alpha}$ because of its simplicity but another method of deconvolution could be used, 
in particular if $\delta_{\alpha}$ is not defined as a Gaussian. 
This algorithm is denoted by ``RL''
throughout.  The RL algorithm uses the following iterative procedure:
\begin{equation}\label{RLeq}
f_{i+1}(\xi) =f_{i}(\xi)\int_\Omega \frac{e^{-\beta A_{\alpha}(\xi^*)}}{\int_\Omega
  \delta_{\alpha}(\hat{\xi}-\xi^*)f_i(\hat{\xi})d\hat{\xi}}
\, \delta_{\alpha}(\xi^*-\xi) \, d\xi^*,
\end{equation} 
where 
$f_0(\xi) = \exp[-\beta A_{\alpha}(\xi)]$, which is given by equation
\eqref{bFE}. To begin the algorithm, $\delta_{\alpha}$ and $f_i$ must
be normalized. The fixed-point iteration~\eqref{RLeq} suggests that $f_n(\xi) \rightarrow
\exp[-\beta A(\xi)]$ as $n \to +\infty$. 
We use 10 iterations in the reported results.

The reference free energy was computed by reweighting a long biased 
trajectory (120~ns) as 
\begin{eqnarray*}
&&A_{\text{ref}}(k,l) =-k_BT
\ln \left( \sum_i \delta[\xi_k-\xi_1(x_i)] \times \right.\\
&& \qquad \qquad \delta[\xi_l-\xi_2(x_i)] \exp[\beta
  V_b(\xi(x_i),\tau)] \Big), 
\end{eqnarray*} where $V_b(\xi,\tau)$ was constructed from 4 ns of simulation with the mollified DOS 
method. The bias was not updated during construction of the reference free energy. 
This produces a
result free from errors associated with the choice of $\alpha$.
The reference profile $A_{\text{ref}}$ is shown in figure \ref{ugly}(A) and in figure
\ref{ugly}(B) we show $A_{\alpha}(\xi,t)$ at 1 ns of sampling with
$\alpha = 5^{\circ}$. The average difference $d(t)$ is shown in figure
\ref{alpcon} for $A_{\alpha}$ with different values of $\alpha$.  To
show how the zero of energy of the bias potential controls the speed
of convergence, in figure \ref{alpcon} we plot Eq.~\eqref{bFE} with
$c=0$ in~\eqref{bias} and we set $c=15k_{B}T$ for the remaining simulations.  
In figure \ref{compcon} we show $d(t)$ 
for Eqs.~\eqref{cle} and~\eqref{dp08t} (ABF methods). 
Results for Eq.~\eqref{metad} (well-tempered metadynamics)
are shown in figure~\ref{metafig}(D).

\subsection{Efficiency of the results as a function of $\alpha$}

For small $\alpha$ the nonlocality of the formulation disappears and
in figure \ref{alpcon} we see slow convergence for $\alpha =
0.8^{\circ}$.  For intermediate values of $\alpha$, nonlocality allows
the bias potential to equilibrate much faster.  For $\alpha
=2^{\circ}$ and $\alpha=5^{\circ}$, $A_{\alpha}$ is a good
approximation of $A$, $d(t)$ falls well under $1$ kcal/mol and we
observe high efficiency.  With the value $\alpha = 10^{\circ}$, $d(t)$
plateaus at roughly $1$ kcal/mol; $\alpha$ is now too large for
$A_{\alpha}$ to be a good approximation of $A$.  After applying the RL
deconvolution to $A_{\alpha=10^{\circ}}$, $d(t)$ drops to match
the accuracy obtained with $\alpha = 2^{\circ}$ or $\alpha = 5^{\circ}$.  The
correspondence between $A_{\alpha}$ and $A$ has deteriorated but not
enough to decelerate the sampling: $A_{\alpha=10^{\circ}}$ is still a 
good biasing potential and $A$ can be recovered with
deconvolution even at very short times.

For large $\alpha$ Eq. \eqref{bwMD} approaches zero, leaving only a
small biasing force to accelerate the dynamics.  To assess whether
$\alpha=20^{\circ}$ is so large as to slow down the sampling, we apply
the RL deconvolution.  The results in figure \ref{alpcon} demonstrate
that $A$ can be recovered to high accuracy for $\alpha=20^{\circ}$ at
long times but that sampling efficiency is affected.

In figure \ref{compcon} we show $d(t)$ for Eqs. \eqref{cle}
and \eqref{dp08t} with a bin size of $1.2^{\circ}$ and also for
Eq. \eqref{dp08t} with a bin width of $3.6^{\circ}$.  If we compare
the time to reach $d(t)=1$ kcal/mol, simulation with Eq. \eqref{bFE}
is roughly three to ten times faster than Eqs. \eqref{cle} and
\eqref{dp08t} for $2 \le \alpha \le 20$ at the bins size of
$1.2^{\circ}$.  For the larger bin size $3.6^{\circ}$, ABF sampling
speed becomes competitive with the mollified density of states
approach but it is impossible to remove the error.  The $3.6^{\circ}$
bin width coincides with the Gaussian half-width of $\delta_{\alpha}$
when $\alpha = 2^{\circ}$.  A larger bin size can enhance sampling
speed for ABF but at a cost in accuracy.  Note that
$\alpha=20^{\circ}$ corresponds to a $\delta_{\alpha}$ with a half
width that spans $33.3^{\circ}$ in one dimension. This is a very large
effective bin width for the accuracy of the results; A similar bin
size with Eqs. \eqref{cle} or \eqref{dp08t} would produce large,
irreparable errors.

In figure \ref{metafig} we show results for the metadynamics simulations. 
We use the values $\Delta T=1800$ K, 
$\omega=0.24$ cal mol$^{-1}$ fs$^{-1}$ and $\tau_G = 120$ fs, as suggested in reference [\citenum{bbp08}]. 
We could not improve the results by choosing different parameters. In panels (A) and (B) of figure \ref{metafig} 
we show the absolute difference between the computed $A_{\alpha=10}$ and the reference $A$ with and without 
deconvolution, respectively. Clearly, the bulk of 
error is due to the missrepresentation of the very negatively curved regions of the free energy and the 
ability to deconvolute $A_{\alpha}$ drastically reduces this error. In panel (C) we show the absolute difference 
between the free energy computed via equation~\eqref{metad} and the reference. We see again that the error is concentrated 
in the regions of large negative curvature but there is not simple and obvious way to reduce these errors with some 
post-process. Panel (D) confirms that the metadynamics promotes extremely rapid sampling but that the long time 
accuracy, especially in strongly curved regions, is limited. 

The results summarized in figures \ref{alpcon}, \ref{compcon} and
\ref{metafig} imply that a wide range of values $2 \le \alpha \le 20$
lead to good efficiency. The ability to use the simple deconvolution
algorithm is a clear advantage of the method. 

\subsection{Choosing $\alpha$ \textit{a priori} }
\label{sec:choice_alpha}

We now discuss how to \textit{a priori} choose $\alpha$ based on some
rough error estimates.  Taking $\xi$ as a scalar, we may expand
$e^{-\beta A(\xi^*)}$ as a Taylor series. 
Eq. \eqref{conv} yields 
\begin{align}\label{rerror}
e^{-\beta (A_{\alpha}(\xi^*)-A(\xi^*))} &\simeq 1+\frac{\alpha^2}{4}
\left[ \left( \frac{A'(\xi^*)}{k_BT}\right)^2 -
  \frac{A''(\xi^*)}{k_BT}\right],
\end{align} where we keep terms up to the second moment of $\delta_{\alpha}$. 
Assuming that $A(\xi)$ is harmonic near the minimum $\xi=q$, the
curvature can be estimated as $A''(q) = k_B T/\sigma^2$ where
$\sigma^2$ is the variance of the reaction coordinate at temperature
$T$.  From Eq.~\eqref{rerror}, 
\[
\exp\big[-\beta(A_{\alpha}(q)-A(q))\big] \simeq 1-\frac{\alpha^2}{4\sigma^2}. 
\]
While the higher order terms and
the regions where $A'\ne 0$ are certainly important to the total
error, this motivates defining $\alpha$ as a function of $\sigma$ if
little is known about the free energy --- we can always calculate
$\sigma$ in the initial state.

We calculate the variance of the reaction coordinates to be about
$\sigma^2 = 340$ degrees squared for both $\Phi$ and $\Psi$.  In terms
of the values of $\alpha$ discussed above, this implies $\sigma /9 \le
\alpha \le \sigma /2$ as a good range for fixing $\alpha$ from
calculation of $\sigma$. 
Of course, different $\alpha$'s may also be used for different
coordinates.

\section{Conclusion}
\label{sec:ccl}

In conclusion, we have developed and tested an efficient ABP
scheme. The nonlocality of $\delta_{\alpha}$ leads to a bias potential
and a bias force that equilibrate rapidly. Shifting the zero of energy
on the bias potential was shown to result in efficient importance
sampling.  The parameter $c$ has influence on only the efficiency of
the importance sampling but not on the limiting error of $A_{\alpha}$.
Because the bias potential is related to a convoluted free energy,
deconvolution can be applied at the end of a simulation to remove all
of the errors associated with the choice of the model parameter
$\alpha$ --- a unique feature and strenght of this approach. 
This is limited only by the extent of sampling and the
spacial discretization.  This scheme easily accommodates the
computation of the free energy surface and free energy gradient in
several dimensions.  We also suggest a simple means of \textit{a
  priori} specifying $\alpha$ and $c$ that should be quite general in
applicability.

\section*{Acknowledgements}

This work is funded by the SIRE project (contract number
ANR-06-CIS-014) of the French national research agency (ANR).

\appendix
\section{A Schematic Algorithm}\label{app:algo}
To help illustrate the simplicity of implementing equation (7) from the text 
\begin{equation}\label{eq:a1}
\frac{\partial A_{\alpha}(\xi^*,n+1)}{\partial \xi^*_j} = -k_BT
\frac{\displaystyle\sum_{i=0}^n \partial_{\xi^*_j}\delta_{\alpha}(\xi(x_i)-\xi^*) e^{\beta V_b(\xi(x_i),i)} 
}{1+\displaystyle\sum_{i=0}^n \delta_{\alpha}(\xi(x_i)-\xi^*)e^{\beta V_b(\xi(x_i),i)} }  
\end{equation} 
for a 2 dimensional 
computation, we give a schematic algorithm here. We first define some array names. Let 
the array named 
``pop($k,l$)'' 
store the population at the $(k,l)$ grid point (this is just the denominator of 
Eq.~\eqref{eq:a1} above), where $k$ corresponds to the bin index of $\xi_1(x_i)$ and $l$ corresponds to the bin index of 
$\xi_2(x_i)$. Let the array named ``dpop($j,k,l$)'' 
hold the derivative of the population along the $\xi_{j=1,2}$ direction at 
the point $(k,l)$. The array ``dpop'' is simply the numerator of Eq.~\eqref{eq:a1} above. 
We use ``dA(k)'' to store the gradient of 
the free energy at the present point $(k,l)$. 
We assume that $\alpha$ has been calculated 
and $c$ has been specified. 
We let $\delta_{\alpha}(\xi) = e^{-|\xi|^2/\alpha^2}$, which amounts to 
ignoring the normalization of the Gaussian functions. 
Lastly, we denote the trajectory in phase space as $x_i$, 
$F(n')$ is the force along the $n'^{th}$ degree of freedom, $d/dn'$ is the derivative with respect to the $n'$ degree of freedom
 and we use $V(x)$ for the potential energy.

First we initialize the arrays. 
\begin{alltt}
\(t=0\), pop\((k,l)=1\) \(\forall k,l\) and 
dpop\((j,k,l)=0\) \(\forall k,l,j\) and \(M=1\),  
\end{alltt} where $M=\max_{k,l}[$pop$(k,l)]$. 
Each time the molecular dynamics forces are computed we must also compute the 
current biasing information. 
Notice that we define everything in terms of 
the ``pop'' and ``dpop'' arrays so that no array is needed for the bias potential and that $M=\max_{k,l}[$pop$(k,l)]$ can be 
updated without looping over the full reaction coordinate domain. 
\begin{widetext}
\begin{alltt}
 ! evaluate free energy gradient at \((k,l)\) for \(j=1,2\) 
\(dA(j) = \)dpop\((j,k,l)/\)pop\((k,l)\)
 ! add bias forces to the existing forces and use a 
 ! ``Ramp function'' R as described in the text
\(R = \)min\((1,\)pop\((k,l)/10)\)
\(F(n')=F(n') + R \sum\sb{j=1}\sp{2} dA(j) d\xi\sb{j}/d{n'}\) 
 ! evaluate the weighting factor \(W\) for updating ``pop'' and ``dpop'' 
\(W = \exp[\beta V\sb{b}] = \exp[\beta c]\)pop\((k,l)/M\) 
 ! update ``pop'' and ``dpop'' on an \(m\) by \(m\) grid 
loop \(k'=k-m/2,k+m/2\)
  loop \(l'=l-m/2,l+m/2\) 
    pop\((k',l')= \)pop\((k',l')+\delta\sb{\alpha}(\xi\sb{1}(x_i) - \xi\sp{*}\sb{1,k'})\delta\sb{\alpha}(\xi\sb{2}(x_i) - \xi\sp{*}\sb{2,l'})W\) 
    if pop\((k',l') > M\) then \(M = \)pop\((k',l')\)  
    loop j=1,2
      dpop\((j,k',l')= \)dpop\((j,k',l')+\partial\sb{\xi\sp{*}\sb{j}}[\delta\sb{\alpha}(\xi\sb{1}(x_i) - \xi\sp{*}\sb{1,k'})\delta\sb{\alpha}(\xi\sb{2}(x_i) - \xi\sp{*}\sb{2,l'})]W\)
\end{alltt}
\end{widetext}
We have defined $k'$ and $l'$ so that ``pop'' and ``dpop'' are updated on an $m$ by $m$ grid as discussed in the text. 
The treatment of $(k',l')$ should reflect whether the domain is assumed to be periodic or not. The 
approximate free energy $A_{\alpha}$ is recovered (up to an additive constant) with $A_{\alpha} =k_BT\ln[\text{pop}(k,l)/M]$. 

The dynamics will now evolve in the presence of the biasing force $dA(j)$, while the arrays ``pop'' and ``dpop'' hold unbiased 
estimates of the population and the derivatives of the population. Notice that the free energy gradient is reduced to a simple 
ratio and the only difficulty lies in the careful treatment of the loops over the grid points $k'$ and $l'$. The often mathematically 
complex computation of the free energy and free energy gradient is reduced to simple bookkeeping. 

\bibliography{xivV2.bbl}

\end{document}